\documentclass[aps,prd,superscriptaddress,12pt,showpacs,notitlepage]{revtex4}
	\usepackage{graphicx}
\usepackage{amsmath}
	\usepackage{amsmath,amssymb,mathrsfs}

\usepackage{epsf}
\usepackage{epsfig}

\newcommand{\be}{\begin{equation}}
\newcommand{\ee}{\end{equation}}
\newcommand{\bea}{\begin{eqnarray}}
\newcommand{\eea}{\end{eqnarray}}

\newcommand{\eqn}{\begin{eqnarray}}
\newcommand{\eqnx}{\end{eqnarray}}

\begin{document}
\title{Rotational-vibrational coupling in the BPS Skyrme model of baryons}

\author{C. Adam}
\affiliation{Departamento de F\'isica de Part\'iculas, Universidad de Santiago de Compostela and Instituto Galego de F\'isica de Altas Enerxias (IGFAE) E-15782 Santiago de Compostela, Spain}
\author{C. Naya}
\affiliation{Departamento de F\'isica de Part\'iculas, Universidad de Santiago de Compostela and Instituto Galego de F\'isica de Altas Enerxias (IGFAE) E-15782 Santiago de Compostela, Spain}
\author{J. Sanchez-Guillen}
\affiliation{Departamento de F\'isica de Part\'iculas, Universidad de Santiago de Compostela and Instituto Galego de F\'isica de Altas Enerxias (IGFAE) E-15782 Santiago de Compostela, Spain}
\author{A. Wereszczynski}
\affiliation{Institute of Physics,  Jagiellonian University,
Reymonta 4, Krak\'{o}w, Poland}


\begin{abstract}
We calculate the rotational-vibrational spectrum in the BPS Skyrme model for the hedgehog skyrmion with baryon number one. The resulting excitation energies for the nucleon and delta Roper resonances are slightly above their experimental values. Together with the fact that in the standard Skyrme model these excitation energies are significantly lower than the experimental ones, this provides strong evidence for the conjecture that the inclusion of the BPS Skyrme model is required for a successful quantitative description of physical properties of baryons and nuclei.
\end{abstract}

\maketitle 
\section{Introduction}
The Skyrme model \cite{skyrme} provides a means to describe the low energy regime of QCD where the baryons and nuclei are topological excitations of the group-valued meson field $U$. In the simplest case, $U\in$ SU(2), and the topological index $Q \in \pi_3$(SU(2)) is identified with the baryon number \cite{thooft}, \cite{nappi}. In its standard version, the model consists of two terms: the sigma model part $L_2$ and the quartic term, i.e., the so-called Skyrme term $L_4$ \cite{manton}. To make the model more realistic, a potential term $L_0$ may be added  \cite{massive}. Also a sextic term $L_6$, the topological current squared, has been advocated as an important ingredient of the full model  \cite{jackson}-\cite{sextic}. In fact, further generalizations which mainly meant the inclusion of further higher derivative terms \cite{higher}, \cite{higher new} as well as new mesonic fields (vector mesons) have been also widely discussed, mainly motivated in the derivative expansion of the large $N_c$ QCD \cite{simic}. Unfortunately, the resulting Lagrangians became more and more complicated, with a rather arbitrary form and growing number of free parameters. As we still are not able to derive the precise form of the model directly from QCD, it is therefore of high importance to better understand the role of different terms of Skyrme type Lagrangians, and to study the possibility to identify a submodel which would give the main and well controlled contributions to some relevant characteristics of baryons and nuclei. The further terms could then be treated as small perturbations about the underlying submodel. 

The crucial point (besides the assumption of the mesonic field $U$ as the proper degrees of freedom, and the solitonic,  i.e., topological nature of the baryons \cite{rho}) in the construction of this model is the almost BPS nature of baryons. Indeed, the energies of the atomic nuclei are with a very good agreement linear in the baryon number. Therefore, such a "hard core" of  Skyrme type effective models should be based on a BPS theory. The BPS Skyrme model is a realization of this idea. It consists of two terms: the sextic part $L_6$ which is the baryon currents squared and a potential $L_0=V$    \cite{BPS}
\begin{equation}
\mathcal{L}_{06}=  -\lambda^2 \pi^4 B_\mu B^\mu - \mu^2 V 
\end{equation}
where the topological (baryonic) current reads
\begin{equation}
B^\mu=\frac{1}{24 \pi^2} Tr(\epsilon^{\mu \nu \rho \sigma}L_\nu L_\rho L_\sigma).
\end{equation}
Here $L_\mu = U^\dagger \partial_\mu U$, $U \in SU(2)$ is the chiral field, and $\mu, \lambda$ are constants. As the model is BPS (static field equations are reducible to a first order ordinary equation) and integrable (in the meaning of the generalized integrability \cite{AFJ}), it offers an {\it analytical} way to better understand many properties of chiral solitons. The BPS Skyrme model provides the required scaling properties of the energy $E$ and radius $R$ with the baryon charge $B$: $E \sim B$ and $R \sim B^{1/3}$. It also has a very large (infinite) group of symmetries \cite{fosco}, which is a subgroup of the SDiff$({\mathbb S}^3)$ on the target space.  Moreover, the static energy integral is invariant under another infinite set of symmetry transformations, the base space SDiff$({\mathbb R}^3)$, that is, exactly the symmetries of an incompressible liquid. Therefore, the model may be treated as a solitonic realization of the liquid droplet limit of nuclear matter. Further arguments for this idea are presented below. We want to remark that an alternative BPS theory has been proposed recently in \cite{poul}.

It is of high importance that the BPS model possesses a well defined Hamiltonian, which, as a consequence, leads to the conventional semiclassical quantization. Indeed, the rotational spectrum of the model for $B=1$ has been found \cite{BPS}, \cite{BoMa}. The corresponding isoscalar/isovector electric/magnetic radii are smaller than in the usual Skyrme model (and closer to the experimental values). This suggests another understanding of the BPS approximation, which can be understood as a limit without usual pion propagation (no pion tails outside nuclei, so the radii are smaller). Indeed, the model does not contain any kinetic term for the pion fields.     
This term should be included as a perturbation. The results  for the binding energies and other properties of nuclei for the first corrections and refinements encourage further work towards better results and deeper understanding.

In the present paper, we thus continue the investigation of the quantum properties of the BPS Skyrme model, beginning with the $B=1$ sector. Specifically, we quantize simultaneously rotational as well as vibrational modes to find the masses of the corresponding Roper resonances,
where collective degrees of freedom should be
important. Again, we find very good results already at the simplest level, as explained in detail in the Conclusions. Roper resonance energies in the standard Skyrme model have been calculated in \cite{Schwes1}, \cite{tarlini}, \cite{zahed} and, for the SU(3) Skyrme model, in \cite{weigel}.
\section{Rotational-vibrational modes}
We start with the BPS Skyrme Lagrangian 
\begin{eqnarray}\label{L06}
&L_{06}&= \int d^3x\left( -\frac{\lambda^2}{24^2}\left[ Tr(\epsilon^{\mu \nu \rho \sigma} L_\nu L_\rho L_\sigma) \right]^2-\mu^2 V  \right) \nonumber \\ && = \int d^3x\left(  -\lambda^2 \pi^4 B_\mu B^\mu - \mu^2 V  \right) \nonumber \\
&& = - (E_0 +E_6) + \lambda^2 \pi^4 \int d^3 x B^i B^i 
 \nonumber \\
&& = - E_{06} + \lambda^2 \pi^4 \int d^3 x B^i B^i 
\end{eqnarray}
\noindent
where $E_0 = \mu^2 \int d^3 x V$ and $E_6 = \pi^4 \lambda^2 \int d^3 x B_0^2 $. By means of the Derrick scaling argument we know that the static energy  is
\begin{equation}
E_{06}=E_0+E_6=2E_0.
\end{equation}
Now we consider the rotations and the scaling transformation simultaneously. Obviously, these two transformations correspond to two physically distinct situations: the rotation (represented by an SU(2) matrix $A$) is a symmetry of the full model while the dilatation transformation $U(x) \rightarrow U(e^\Lambda x) $ does change the action. Then we semiclassically quantize both modes in the usual way. First we assume that the transformation parameters are time-dependent and then we promote them to some quantum mechanical variables which are subject to quantization. 
\\
The breathing and rotational motions are given by the following transformation 
\begin{equation}
U(x) \rightarrow U'(x,t)=A(t)U_0(xe^{\Lambda (t)}) A^{-1}(t)
\end{equation}
Obviously, the static part of the Lagrangian changes only due to the scaling 
\begin{equation}
-E_{06} \rightarrow -(e^{-3\Lambda(t)} E_0 + e^{3\Lambda(t)} E_6) .
\end{equation}
To compute the remaining time dependent contribution, we need to know how the time component changes under the considered transformations,
\begin{eqnarray}
&L_0(x)& \rightarrow L'_0(x')=A U^\dagger_0(xe^{\Lambda (t)}) A^{-1} \partial_0 \left( A U_0(xe^{\Lambda (t)}) A^{-1} \right) \\ & & = A U_0^\dagger (x') \partial_{m'} U_0(x') A^{-1} \frac{dx'^m}{dt} \\ & & +AU_0^\dagger(x') A^{-1}\dot{A} U_0(x') A^{-1} +A U_0^\dagger(x') A^{-1} A U_0(x') \dot{A^{-1}}\\ & &=   A U_0^\dagger (x') \partial_{m'} U_0(x') x'^{m} A^{-1} \dot{\Lambda} \\ && +AU_0^\dagger(x') A^{-1}\dot{A} U_0(x') A^{-1} -\dot{A} A^{-1} \\ && = A L_{m'} x'^{m}A^{-1} \dot{\Lambda}  +AU_0^\dagger(x') A^{-1}\dot{A} U_0(x') A^{-1} -\dot{A} A^{-1} 
\end{eqnarray}
where $x' = xe^{\Lambda (t)}$. 
Further
\begin{eqnarray}
&L_j(x)& \rightarrow L'_j(x')=A U^\dagger_0 (xe^{\Lambda (t)}) A^{-1} \partial_j \left( A U_0(xe^{\Lambda (t)}) A^{-1} \right) \\ & & = A U_0^\dagger (x') \partial_{j'} U_0(x') A^{-1} e^{\Lambda} .
\end{eqnarray}
Hence, the space component of the baryon current reads
\begin{eqnarray}
&B^i(x) & \rightarrow \frac{3}{24\pi^2} \epsilon^{i0jk} Tr \left( L_0'(x')L'_j(x')L'_k(x')\right) \\ && =
\frac{3}{24\pi^2} e^{2\Lambda(t)}  \epsilon^{i0jk}  \left( Tr (L_{m'} x'^mL_{j'}L_{k'}) \dot{\Lambda} \right. \\ && \left.+Tr \left( U^\dagger_0(x') [A^\dagger \dot{A},U_0(x')] L_{j'}L_{k'}\right)\right) .
\end{eqnarray}
Now, it is useful to notice that 
\begin{equation}
 \epsilon^{i0jk}  Tr (L_{m} x^mL_{j}L_{k}) = - Tr(L_1,[L_2,L_3]) x^i =  \frac{24\pi^2}{3} B_0 x^i .
\end{equation}
At this stage, we may treat the parameters of the transformations as quantum mechanical coordinates, with the usual assumption for the rotation  
$A^\dagger \dot{A}=\frac{i}{2}\vec{\Omega} \cdot \vec{\tau}$, where $\vec{\tau}$ are the Pauli matrices and $\vec \Omega$ are the angular velocities. Then, 
\begin{equation}
U^\dagger_0(x) [A^\dagger \dot{A},U_0(x)]= \Omega_i T_i, \;\;\;\;\; T_i=\frac{i}{2}U_0^\dagger [\tau_i, U_0(x)]
\end{equation}
and we get 
\begin{eqnarray}
&& \lambda^2 \pi^4 \int d^3x B_i^2 \rightarrow \\ & &
\lambda^2 \frac{3^2}{24^2} \int d^3 x e^{\Lambda} \left( \frac{24\pi^2}{3} B_0x^i \dot{\Lambda} +\epsilon^{ijk} \Omega_a Tr(T_aL_jL_k) \right)^2 \\ & & = \lambda^2 \pi^4 e^{\Lambda(t)} \dot{\Lambda}^2 \int d^3 x B_0^2 r^2 \\ & &
+ 2\lambda^2 \pi^2 \frac{3}{24} e^{\Lambda(t)} \dot{\Lambda} \Omega_a  \int d^3 x B_0 \epsilon^{ijk} x^i Tr (T_aL_jL_k)  \\ & &
+ \lambda^2 \frac{3^3}{24^2} e^{\Lambda(t)} \Omega_a \Omega_b \int d^3 x  \epsilon^{ijk} \epsilon^{irs} Tr (T_a L_jL_k) Tr (T_b L_rL_s) .
\end{eqnarray}
One can check that the second integral vanishes while the third provides the moment of inertia tensor ${\cal U}_{ab}$, 
\begin{equation}
\int d^3 x  \epsilon^{ijk} \epsilon^{irs} Tr (T_a L_jL_k) Tr (T_b L_rL_s) = \frac{1}{2} \frac{24^2}{3^2 \lambda^2} {\cal U}_{ab} .
\end{equation}
Hence,
\begin{equation}
\lambda^2 \pi^4 \int d^3x B_i^2 \rightarrow \lambda^2 \pi^4 e^{\Lambda(t)} \dot{\Lambda}^2 \int d^3 x B_0^2 r^2 + \frac{1}{2} e^{\Lambda(t)} \Omega_a {\cal U}_{ab} \Omega_b .
\end{equation}
Finally, the full Lagrangian reads
\begin{equation}
L_{v+r}= e^{\Lambda(t)} \dot{\Lambda}^2 Q_6 -(e^{-3\Lambda(t)} E_0 + e^{3\Lambda(t)} E_6)+ \frac{1}{2} e^{\Lambda(t)} \Omega_a {\cal U}_{ab} \Omega_b
\end{equation}
where 
\begin{equation}
Q_6=  \lambda^2 \pi^4  \int d^3 x B_0^2 r^2 .
\end{equation}
The Hamiltonian is given by the following formula 
\begin{equation}
\mathcal{H}_{r+v}=e^{-\Lambda} \frac{p^2}{4Q_6} + (e^{-3\Lambda(t)} E_0 + e^{3\Lambda(t)} E_6)+\frac{1}{2} e^{-\Lambda(t)} \left( \frac{L_1^2}{{\cal U}_{11}} + \frac{L_2^2}{{\cal U}_{22}}  +\frac{L_3^2}{{\cal U}_{33}} \right)
\end{equation}
where $p$ is the momentum conjugate to $\Lambda$, and $\vec L$ is the body-fixed angular momentum. Using the explicit charge $B=1$ solution of the BPS Skyrme model \cite{BPS} (with the old Skyrme potential $V=\frac{1}{2} \mbox{Tr} (1-U)$) we find that 
\begin{eqnarray}
&Q_6& = 16 \lambda^2 \int_0^{R_0} dr r^2 4\pi \frac{1}{R_0^6} \left(1-\frac{r^2}{R_0^2}\right)r^2 \\ & &
=16 \cdot 4\pi \lambda^2 R_0^{-1}  \int_0^{1} d\tilde{r} \tilde{r}^4\left(1-\tilde{r}^2\right) \\ & &
= 16 \cdot 4\pi \lambda^2 R_0^{-1} \cdot \frac{2}{35} 
= \frac{32}{35} 4\pi \lambda^2 \left( \frac{\sqrt{2}}{4} \right)^{1/3} \left( \frac{\mu}{\lambda} \right)^{1/3} 
\end{eqnarray}
where 
\begin{equation}
R_0 = \left( \frac{\sqrt{2} \mu}{4\lambda} \right)^{-1/3}
\end{equation}
is the size of the BPS Skyrmion (compacton). Further, all off-diagonal components of the inertia tensor vanish while diagonal ones take the same non-zero value
\begin{equation}
{\cal U}\equiv {\cal U}_{11}={\cal U}_{22}={\cal U}_{33}=\frac{2^8\sqrt{2}\pi}{15\cdot 7} \lambda \mu \left( \frac{\lambda}{\mu} \right)^{2/3} .
\end{equation}
Finally, the energy is \cite{BPS}
\begin{equation}
E_{06}=\frac{64\sqrt{2}\pi}{15} \mu \lambda .
\end{equation}
Then,
\begin{equation}
\mathcal{H}_{r+v}=e^{-\Lambda} \frac{p^2}{4Q_6} + (e^{-3\Lambda(t)} E_0 + e^{3\Lambda(t)} E_6)+\frac{1}{2} \frac{e^{-\Lambda(t)}}{{\cal U}} \vec{L}^2 .
\end{equation}
Now we compute the expectation value of the Hamiltonian in a state with fixed spin $|j\rangle$ (we use $\vec L^2 = \vec J^2$, where $\vec J$ is the space-fixed angular momentum operator; remember that
 spin equals isospin for the hedgehog skyrmion with $B=1$) 
\begin{equation}
H_{r+v}=e^{-\Lambda} \frac{p^2}{4Q_6} + (e^{-3\Lambda(t)} E_0 + e^{3\Lambda(t)} E_6)+\frac{1}{2} \frac{e^{-\Lambda(t)}}{{\cal U}} j(j+1) \hbar^2 .
\end{equation}
In each fixed $j$ sector the ground state for $\Lambda$ is given by the condition $\partial {\cal V}(\Lambda) / \partial \Lambda =0$, where the quantum mechanical potential is
\begin{equation}
{\cal V}=e^{-3\Lambda(t)} E_0 + e^{3\Lambda(t)} E_6+\frac{1}{2} \frac{e^{-\Lambda(t)}}{{\cal U}} j(j+1) \hbar^2 .
\end{equation}
Then, $\Lambda_0(j)$ is a solution of the following equation
\begin{equation}
-3e^{-3\Lambda_0(j)} E_0 + 3e^{3\Lambda_0(j)} E_6 - \frac{1}{2} \frac{e^{-\Lambda_0(j)}}{{\cal U}} j(j+1) \hbar^2=0 .
\end{equation}
Obviously, the scaling transformation is not a symmetry, therefore, in the first approximation, it is enough to keep only the quadratic terms in a perturbation about the ground state, $\Lambda=\Lambda_0+\epsilon$. This gives the harmonic oscillator Hamiltonian for $\epsilon$
\begin{equation}
H_{r+v}=e^{-\Lambda_0(j)} \frac{p^2}{4Q_6} + 2E_0(-e^{-3\Lambda_0(j)} +2 e^{3\Lambda_0(j)})+\epsilon^2 3E_0(e^{-3\Lambda_0(j)} + 2e^{3\Lambda_0(j)})
\end{equation}
where we used that $E_0=E_6$. Then, the energy spectrum takes the form
\begin{equation}
E_{n,j}=2E_0(-e^{-3\Lambda_0(j)} +2e^{3\Lambda_0(j)}) +\left(n+\frac{1}{2} \right) \left( \frac{3E_0(e^{-4\Lambda_0(j)} +2 e^{2\Lambda_0(j)})}{2Q_6} \right)^{1/2} \hbar .
\end{equation}
Now we may fit the values of the parameters to the masses of the nucleon $E_{0,\frac{1}{2}}=M_N=938.9 $ MeV and the $\Delta$ resonance $E_{0,\frac{3}{2}}=M_\Delta=1232$ MeV. This leads to $\mu \lambda = 30.06$ MeV and $\lambda^2 \left( \frac{\mu}{\lambda} \right)^{1/3}=17.79$ Mev fm$^2$. The values of the excitation masses for the nucleon Roper and $\Delta$ Roper resonances are given in Table I. They are compared with values obtained using the same method in the massless  Skyrme model (for excitation energies found by other methods see \cite{meissner}).

\begin{table} \label{tab}
\begin{center}
\begin{tabular}{|c|c|c|c|c|}
\hline
  & BPS Skyrme  & Skyrme \cite{zahed} & Skyrme \cite{tarlini} & experiment \cite{data} \\
\hline 
$N$(1440) & 588 & 390  & 388 &502 $\pm$ 20 \\
$\Delta$ (1600) &  593 &  290 & 292 &368 $\pm$ 100 \\ 
\hline 
\end{tabular}   
\caption{The excitation energies of the nucleon Roper resonance and the $\Delta$ Roper resonance with respect to the nucleon and $\Delta$ mass respectively. All numbers are in MeV} \label{table}
\end{center}
\end{table}

\section{Conclusions}
The main result of the present paper is the observation that the sextic term $B_\mu^2$ rather significantly increases the masses of the vibrational modes. If we stay within the BPS Skyrme model, which contains only the sextic and the potential terms, the masses are slightly too high in comparison to the experimental data. Quite surprisingly, the BPS Skyrme model, although simple and integrable, provides better agreement than the usual Skyrme model, built out of the sigma model term and the Skyrme term, which gives excitation masses which are lower than the true values. The fact that the BPS model gives slightly too high
resonance energies, whereas the standard Skyrme model gives quite too low
energies, points towards the conjecture that a complete model with all
four terms, but where the BPS model terms give the main contribution to
the solitons energies, should give rather good resonance energies.
All that may indicate that the sextic term i.e., the topological current squared, plays a prominent role in the application of Skyrme theories to low-energy QCD, and its inclusion seems therefore to be rather unavoidable.   

There are several comments in order. Firstly, the higher mass provided by the BPS Skyrme model can be understood as the result of a higher incompressibility of the model, as compared to the usual Skyrme model. Indeed, simple scaling arguments indicate that the incompressibility
modulus for the BPS Skyrme model is significantly bigger than in the
standard massless Skyrme model. Then, obviously, it costs much more energy to excite a breathing mode. This, together with the known fact that the static energy integral is invariant under the volume preserving diffeomorphisms (VPDs) on base space, further supports the interpretation of the BPS Skyrme model as an effective solitonic model of the nuclear matter in the incompressible liquid approximation.  

Secondly, if the nucleon Roper resonance is, instead, computed without taking into account the coupling between the vibrational and the rotational degrees of freedom, then it is slightly heavier, with a mass of about $1600$ MeV, instead of the obtained $1527$ MeV. So, qualitatively the vibrational and vibrational-rotational spectra look similar. This is somehow different to the Skyrme model, where the strong rotational-vibrational coupling does change the spectrum. Otherwise, the first breathing mode of the nucleon is too soft with its mass below the first rotational excitation \cite{tarlini}.  

Thirdly, in the present computation we have taken into account only the rotational zero modes and  ignored all other zero modes, whose number is, in fact, infinite as the symmetry group is the VPD group.  The main reason for that (apart from the obvious difficulties in quantizing the full set of VPD transformations) is that finally one has to add at least the sigma model term into the full effective chiral Lagrangian to make the model applicable also in the near vacuum regime. Then, the VPD symmetry will be explicitly broken and the corresponding zero modes will disappear. However, the main message we have learnt from the BPS Skyrme model is that such a symmetry breaking should happen in a rather soft way. 
Hence, the resulting excitations (quadrupole as well as higher
momenta) should possess relatively small energies, which again is
confirmed in experiments (and observed, e.g., in  the $B=4$ skyrmion, where
the first nine vibrational modes correspond to deformations which
involve incompressible flow \cite{barnes}, see also \cite{skyrme-vib}). 
Undoubtedly this issue requires further detailed studies.      

Finally, to some extent, the form of the potential term is rather arbitrary in the BPS model. As the model does not contain any propagation term for the pion field, the potential does not have any direct interpretation as a mass term. Thus, in principle there is no strong restriction on it. Here we have used the so-called old Skyrme potential, but one may consider more general functions to get better agreement with the experimental data.

\section*{Acknowledgement}
The authors acknowledge financial support from the Ministry of Education, Culture and Sports, Spain (grant FPA2008-01177), 
the Xunta de Galicia (grant INCITE09.296.035PR and
Conselleria de Educacion), the
Spanish Consolider-Ingenio 2010 Programme CPAN (CSD2007-00042), and FEDER. 
CN thanks the Spanish
Ministery of
Education, Culture and Sports for financial support (grant FPU AP2010-5772).
Further, AW was supported by polish NCN grant 2011/01/B/ST2/00464. AW thanks Nick Manton for discussions on the issue of incompressibility in the Skyrme models.

\end{document}